\documentstyle[jkas]{article}

\def\kms{km s$^{-1}$\ }
\def\kmss{km s$^{-1}$} 
\def\simlt{\lower.5ex\hbox{$\; \buildrel < \over \sim \;$}}
\def\simgt{\lower.5ex\hbox{$\; \buildrel > \over \sim \;$}}
\def\twoco{$^{12}$CO\ }
\def\threeco{$^{13}$CO\ }

\runningauthor {J.-W. LEE ET AL.} 
\runningtitle{CO Observations of G54.1+0.3}

\month{October} \year{2012} \volume{45} \issueno{5}
\date{Received August 3, 2012; Accepted September 8, 2012}

\begin{document}
\title{CO J=2-1 LINE OBSERVATIONS TOWARD THE SUPERNOVA REMNANT G54.1+0.3} 
\author{Jung-Won Lee$^{1}$, Bon-Chul Koo$^{2}$, and Jeong-Eun Lee$^{3}$   } 
\address{$^1$ Korea Astronomy and Space Science Institute,
, Daejeon 305-348, Korea\\
 {\it E-mail : jwl@kasi.re.kr}}
\address{$^2$ Department of Physics and Astronomy, Seoul National University,
Seoul 151-742, Korea\\ {\it E-mail : koo@astro.snu.ac.kr}}
\address{$^3$ Department of Astronomy and Space Science, Kyung Hee University,
Yongin 446-701, Korea\\ {\it E-mail :  jeongeun.lee@khu.ac.kr}}
\address{\normalsize{\it (Received August XX, 2012; Accepted September XX, 2012)}}
\offprints{J.-W. Lee}
\abstract{We present \twoco $J=2$--1 line observations of G54.1+0.3, a composite 
supernova remnant with a mid-infrared (MIR) loop surrounding the 
central pulsar wind nebula (PWN).
We mapped an area of $12'\times 9'$ around the PWN 
and its associated MIR loop. 
We confirm two velocity components that had been proposed to 
be possibly interacting with the PWN/MIR-loop; 
the +53 \kms cloud that appears in contact with the eastern boundary of the PWN
and the +23 \kms cloud that has CO emission coincident with the MIR loop.
We have not found a direct evidence for the interaction in either of 
these clouds. Instead, we detected an 5$'$-long arc-like cloud at +15--+23 \kms with a systematic 
velocity gradient of $\sim$3 km s$^{-1}$ arcmin$^{-1}$ and broad-line emitting CO gas having widths (FWHM) of $\simlt7$ \kms in the western interior of the supernova remnant. 
We discuss their association with the supernova remnant. 
 }

\maketitle

\section{INTRODUCTION}

G54.1+0.3 is a young, core-collapse supernova remnant (SNR). 
The remnant has a central PWN
with a 136 ms radio/X-ray pulsar 
(PSR J1930+1852) at the center of the nebula 
\citep{lu02}. The characteristic age of the pulsar is 2,900 yr.
The remnant had been known as a Plerion or Crab-like SNR 
of $120''\times 80''$ size because of the central PWN,
but recently, a faint radio emission with a diameter
of $\sim 10'$ surrounding the PWN has been detected by
\cite{lang10}, which was proposed to be the SNR shell driven by 
supernova (SN) ejecta of G54.1+0.3.
Another evidence for the SNR shell was provided by 
\citet{bocchino10}, who detected diffuse, thermal X-ray 
emission filling the radio shell.
Therefore, G54.1+0.3 is now one of the composite SNRs where we 
observe a central PWN surrounded by an extended SNR shell.
The estimated distances to the SNR range from 6 to 9 kpc (see \S~4).

In G54.1+0.3, \citet{koo08} detected a MIR loop surrounding the 
central PWN. The loop is partially complete, and 
is elongated along the northwest-southeast direction. 
It has an extent of $\sim 105''\times 54''$ and surrounds 
the southern part of the PWN.
There are eleven stellar sources with strong
MIR excesses embedded in the loop. \citet{koo08} showed that they 
are OB-type stars and, based on their mid/far-infrared excesses and 
spatial confinement in a loop-like structure
surrounding the PWN, proposed that they are 
young stellar objects whose
formation was triggered by the progenitor of the SNR. 
But later \citet{temim10} carried out Spitzer spectroscopy of the MIR shell,
and detected dust emission with a bump at 21 $\mu$m which was
similar to the emission feature of freshly-formed SN dusts detected in the young 
SNR Cassiopeia A. This, together with broad
ionic emission lines, led \citet{temim10} to propose a scenario
where the stellar objects are the members of a cluster
that the progenitor belonged, and the IR emission comes from SN dusts.
So there are two scenarios for the nature of the MIR loop and 
the stellar objects.

In this paper, we present \twoco $J$=2--1 emission line
observations of the IR loop and the surrounding area.
The molecular environment is expected to be quite different for 
the two scenarios, so that the observation will be useful 
in understanding the nature of the IR loop and the stellar sources. 
\citet{koo08} found that there was a faint $^{13}$CO $J=1$--0 emission
coincident with the IR loop in
the Boston-University-Five College Radio Astronomy Observatory 
Galactic Ring Survey \citep{Jackson06}. They proposed that this 
molecular cloud could be associated with the SNR.
On the other hand, \citet{leahy08}, using the same survey data, found that 
there was a CO cloud that appears to be blocking the eastern boundary of 
the PWN and proposed that this cloud is interacting with 
the SNR. Both propositions are based on 
circumstantial evidence and the association needs further investigation.
Recently $\gamma$-ray emission has been detected toward the PWN, 
which could be partly due to the decay of neutral pions 
produced in the interactions of pulsar-accelerated  
nuclei with one of these molecular clouds \citep{li10, acciari10}.
Our observation, however, has not detected 
a direct evidence for the interaction of the PWN with either cloud.
We present our observation in next sections.

\begin{figure}[!t]
\centering 
\epsfxsize=0.45\textwidth 
\epsfbox{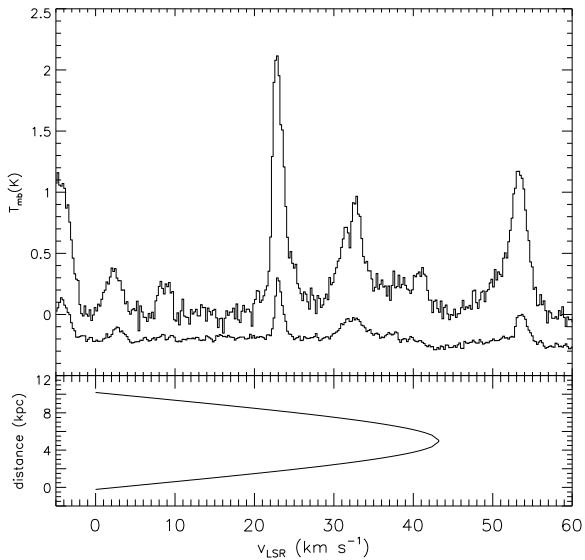} 
\caption{
(top) Average $^{12}$CO {\it J} = 2--1 spectrum  
over the observed  12$'\times9'$ area. Overlaid at
temperature offset of $-0.25$ K is 
the corresponding $^{13}$CO {\it J} = 1--0 spectrum obtained 
from the FCRAO Galactic Ring Survey.
(bottom) Relation between the distance and LSR velocity based on the 
rotation curve of \citet{brand93}.}
\label{avg_spectrum}
\end{figure}

\section{OBSERVATIONS}

$^{12}$CO $J$=2-1 mapping observations of G54.1+0.3 were done in February and March, 2008 using a 230 GHz SIS mixer receiver \citep{jwl08} newly installed on the 6-m telescope at Seoul Radio Astronomy Observatory (SRAO). The new 210--265 GHz band receiver shows quantum-limited noise performance. It employs a RF and an IF quadrature hybrid for sideband rejection and supports dual-polarization observation. At the time of initial operation, only single polarization mode was supported. The observation of G54.1+0.3 was the first science observation after commissioning runs of the new receiver system. 
An autocorrelation spectrometer was configured to cover 100 MHz with 2048 channels, which corresponds to 130 \kms velocity coverage. 
The system temperature was 200--400 K and the typical $\it{rms}$ noise level of the spectra is $\sim$ 0.13 K at 0.21 \kms velocity resolution. 
All the spectra
presented in this paper have been re-gridded to 
0.21\kms velocity resolution to match the $^{13}$CO $J$=1-0 data from the Galactic Ring Survey \citep{Jackson06}. 
The main beam efficiency of the telescope $\eta_{mb}$, measured using Mars, is about 0.51 at 230 GHz. We are using $\rm T_{mb} = T^{*}_{A} /\eta_{mb}$ for temperature scale throughout this paper, if not otherwise stated.   
The mapped area is 12$'\times9'$ around the remnant G54.1+0.3, centered at ($\alpha,\delta$)$_{2000}=$(19$^{h}$30$^{m}$37.5$^{s}$,+18$^{\circ}$52$'$44$''$). The spectra were obtained at a grid spacing of 30$''$.  
The beam size of the SRAO telescope at 230 GHz is 48$''$.
In the Galactic Ring Survey, \threeco data were 
obtained at 22$''$ grid spacing and beam width of 46$''$. 
We have re-sampled the  \threeco data cube with 30$''$ grid spacing using the Miriad data analysis package.   
 
\begin{figure}[!t]
\centering
\epsfxsize=0.45\textwidth \epsfbox{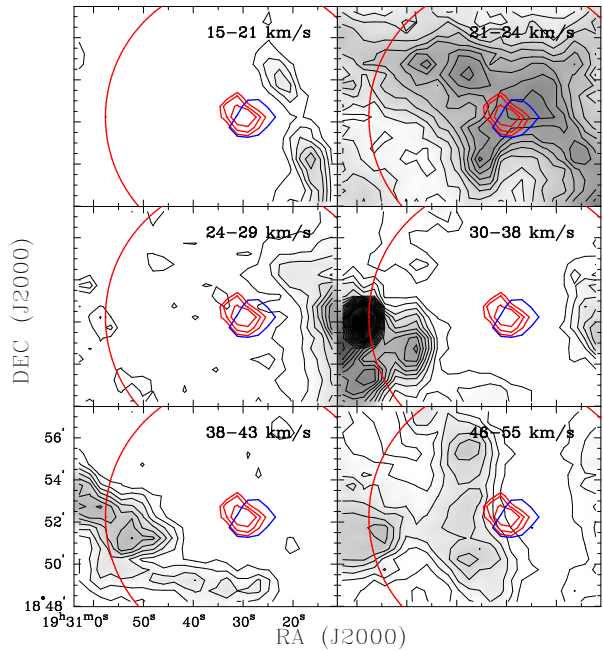} 
\caption{
\twoco $J$=2--1 integrated intensity maps showing individual velocity components
at $v_{\rm LSR}\ge 15$ \kmss. 
Integrated velocity ranges are marked on individual channel maps.
Contour levels increase from 0.2 K km s$^{-1}$ in 0.2 K km s$^{-1}$ steps. 
The red contours show the 21-cm continuum brightness distribution of
the central PWN of G54.1+0.3,
while the blue contour shows the shape of the MIR shell at 70 $\mu$m.
The large red circle
of radius $6.'5$ represents the
outer boundary of the SNR blast wave \citep{lang10,bocchino10}.
The pulsar PSR J1930+1852 is located at  ($\alpha,\delta$)$_{J2000}=$(19$^{h}$30$^{m}$30$^{s}.13$,+18$^{\circ}$52$^{'}$14$^{''}.1).$}
\label{overall_channel_map}
\end{figure}

\begin{figure*}[!ht]
\centering \epsfxsize=0.75\textwidth 
\epsfbox{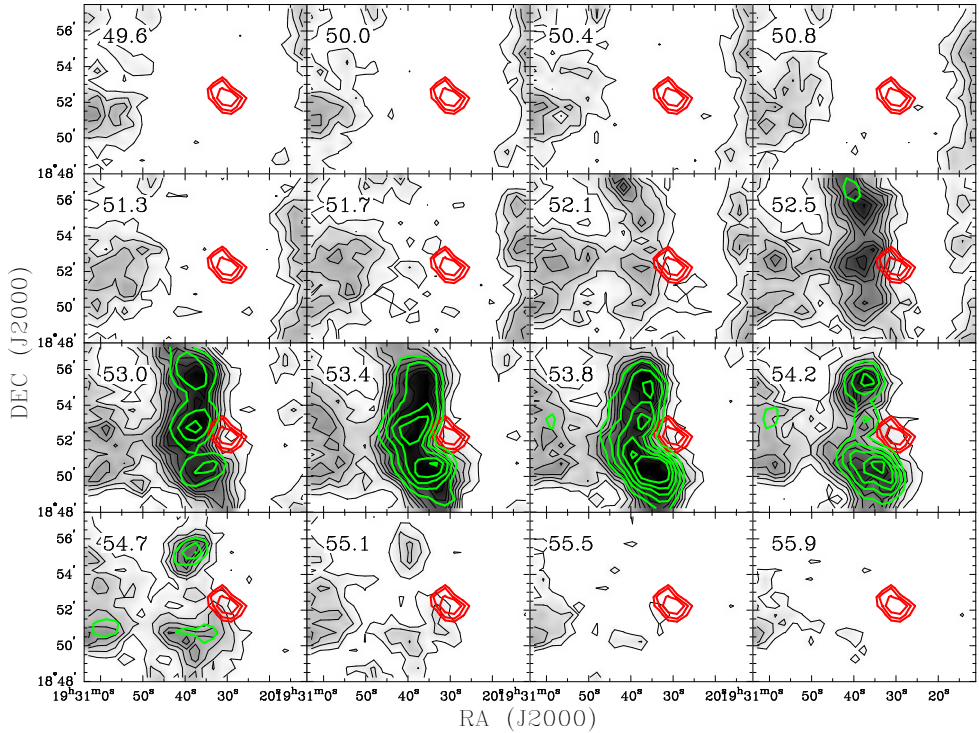} 
\caption{
\twoco $J$=2--1 channel maps of the +53 \kms cloud. The contours have 0.4 K steps from 0.4 K. Again the red contours represent the PWN.
The green contours show 
the \threeco distribution leveled from 0.4 K with 0.2 K steps. 
}

\label{53_channel}
\end{figure*}

\section{RESULTS}
\subsection{Overall Distribution of CO Gas}

Figure ~\ref{avg_spectrum} shows the average spectrum of $^{12}$CO $J$ = 2--1 over the observed region with corresponding $^{13}$CO {\it J} = 1--0 spectrum obtained from the Galactic Ring Survey. Between $v_{LSR}$ = 0 and 60 \kms distinct 
spectral components appear  
at 2, 8, 23, 32, 40 and 53 \kms in the $^{12}$CO spectrum. 
In the bottom frame of the figure, we show the relation between the distance
and LSR velocity according to the rotation curve of \citep{brand93}
which is close to the flat rotation curve with $R_{\odot}$ = 8.5 kpc and 
$\Theta_{\odot}$ = 220 \kmss. Note that there are two distances corresponding to a given 
positive LSR velocity, i.e., +23 \kms component could be either at 1.8 kpc or at 8.2 kpc. 
The maximum LSR velocity at this Galactic longitude (54.$^{\circ}1$) 
is about 43 \kms according to this model.
But there are observations suggesting that the maximum 
rotation velocity toward this Galactic longitude is somewhat higher 
than 220 \kms \citep[see][and references therein]{leahy08}, so that 
the +53 \kms component might be near the tangential point which is 
6.9 kpc assuming the above Galactocentric distance and Galactic 
rotational velocity at the Sun.

The integrated intensity maps at every 5--9 \kms from 15 to 55 \kms 
are presented in Figure ~\ref{overall_channel_map}. 
At $v_{LSR}$ = 15--21 \kmss, there is a faint arc-like cloud 
elongated along the north-south direction in the 
western side of the PWN. As will be shown below, this arc-like 
cloud (hereafter $`$Arc Cloud') has a large line width and 
systematic velocity gradient along the cloud.
At $v_{LSR}$ = 21--24 \kmss, there is strong, extended emission 
superposed on the PWN. 
It has a rather complex morphology with several peaks. 
Between $v_{LSR}$ = 24 and 43 \kmss, CO clouds appear 
in the surrounding area, i.e., in the western and southeastern 
areas at 24--29 \kms and 30--43 \kmss, respectively.
They are well separated from the PWN/IR-loop and extend 
beyond the SNR boundary, so that they are 
not considered to be possibly associated with the PWN (see \S3.2).
At 45--60 \kmss, there is another thick, arc-like cloud that appears in 
contact with the eastern boundary of the PWN. This is the cloud 
that was proposed to be interacting with the PWN by \citet{leahy08}.
In the below, we first investigate the properties of the +53 \kms component
and then the +23 \kms component.

\subsection{+53 \kms Cloud}

\begin{figure}[!t]
\centering \epsfxsize=0.4\textwidth
\epsfbox{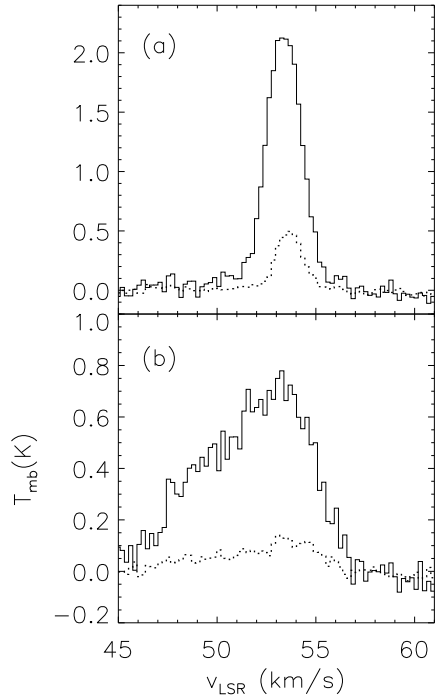} 
\caption{
Average \twoco $J=$2--1 (solid) and \threeco $J=$1--0 spectra (dotted) of (a) the 53 \kms cloud and (b) the 
diffuse broad line emission in the eastern area of the cloud.}
\label{53_broad}
\end{figure}

Figure \ref{53_channel} shows channel maps of the +53 \kms cloud
where we overplot the contours of the 
$^{13}$CO 1--0 emission at corresponding velocities. 
The cloud is composed of three clumps of a few arc-minute extents;
the southern brightest one centered at 53.9 \kmss,
a faint middle one at 53.5 \kmss, and the 
northern one of moderate brightness at 54.3 \kmss. 
The northern one is superposed with another filamentary cloud 
in the northern area at slightly lower velocity (52.6 \kmss).
As pointed out by \citet{leahy08}, the PWN nebula is located 
in the middle where the cloud appears to kink.
But we do not see any indication of the interaction 
in the line profiles of this cloud (Figure \ref{53_broad}). 
The ratio $^{12}$CO(2--1)/$^{13}$CO(1--0) ($\equiv R_{21/10}$)$\approx
5$ which is a typical value for a calm molecular cloud \citep[e.g., see][]{sakamoto94}.

An interesting feature is the weak ($\sim 1$ K) 
emission between 45 and 55 \kms 
in the {\it eastern} area of the +53 \kms cloud. As shown in Figure \ref{53_broad},
the CO line profiles are broad ($\sim 10$ \kmss) although 
there seems to be more than one component.  
It extends beyond the eastern boundary of the our mapping area, so that 
we could not see its full extent. The association of this 
broad emission with the +53 \kms cloud is not clear because the two clouds are connected by faint emission in the middle. 
Since the radio boundary
of the G54.1+0.3 SNR appears quite circular and the
broad emission extends beyond the boundary, we 
consider that this component is not related to the SNR.
(The radio filaments outside
the SNR shell shown in Figure 4 of Lang et al. (2010)
are believed to be belong to a much larger-scale 
structure. See Figures 1--2 of Leahy et al. (2008).)
Instead there is a star-forming infrared filament in the Spitzer MIPS
24 $\mu$m image that is spatially coincident with the broad CO emission as
well as  the 30--38 \kms CO emission in Figure \ref{overall_channel_map}, and we
suspect that the broad CO emission is associated with this filament.

\subsection{+23 \kms Clouds}

\begin{figure*}[!t]
\centering
\epsfxsize=0.8\textwidth \epsfbox{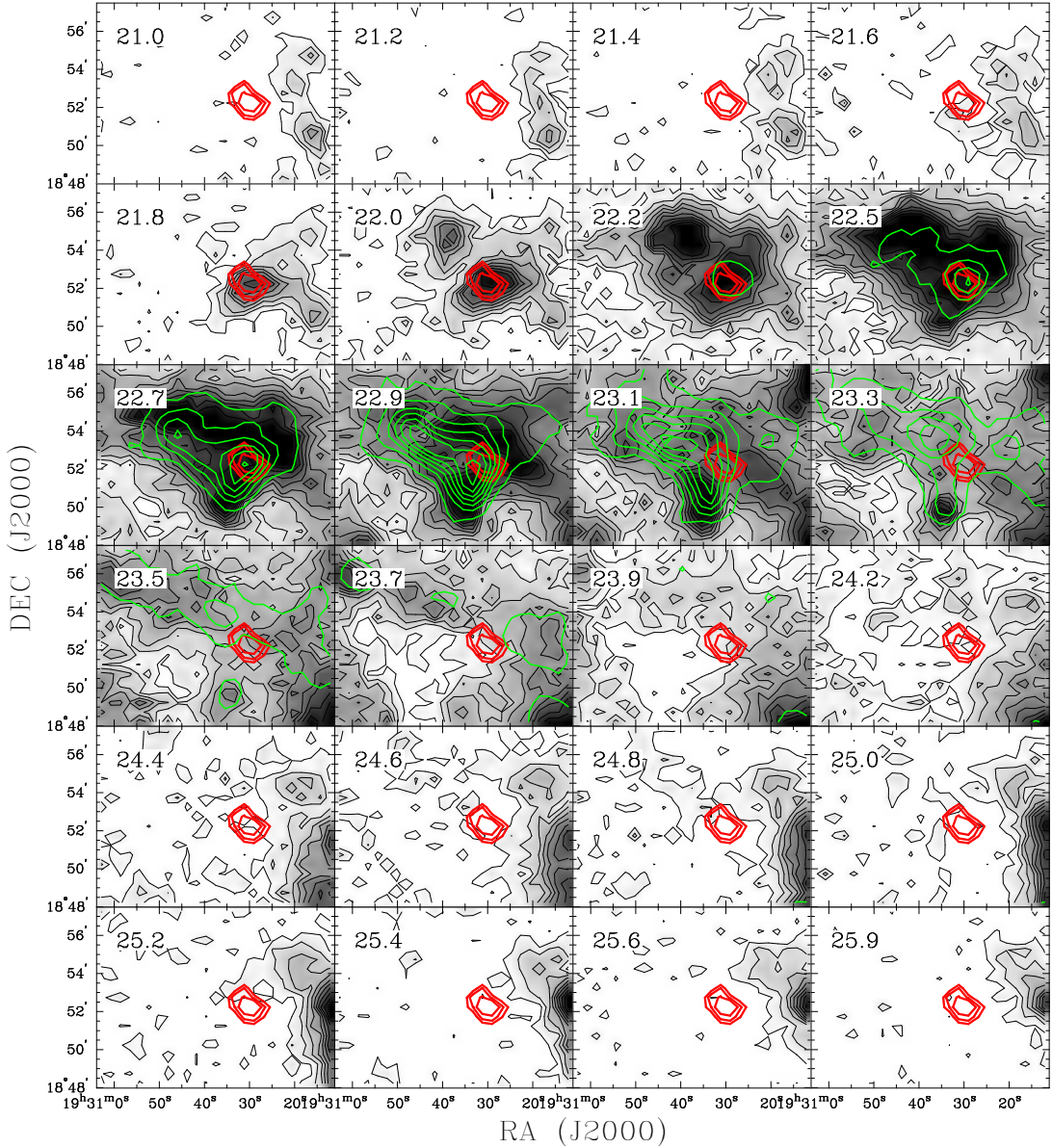} 
\caption{
Same as Fig.~\ref{53_channel}, but for the +23 \kms cloud. 
\twoco contour levels increase from 0.4 K in 0.4 K steps. 
}
\label{23_channel}
\end{figure*}

\begin{figure}[!t]
\epsfxsize=0.5\textwidth
\centering \epsfbox{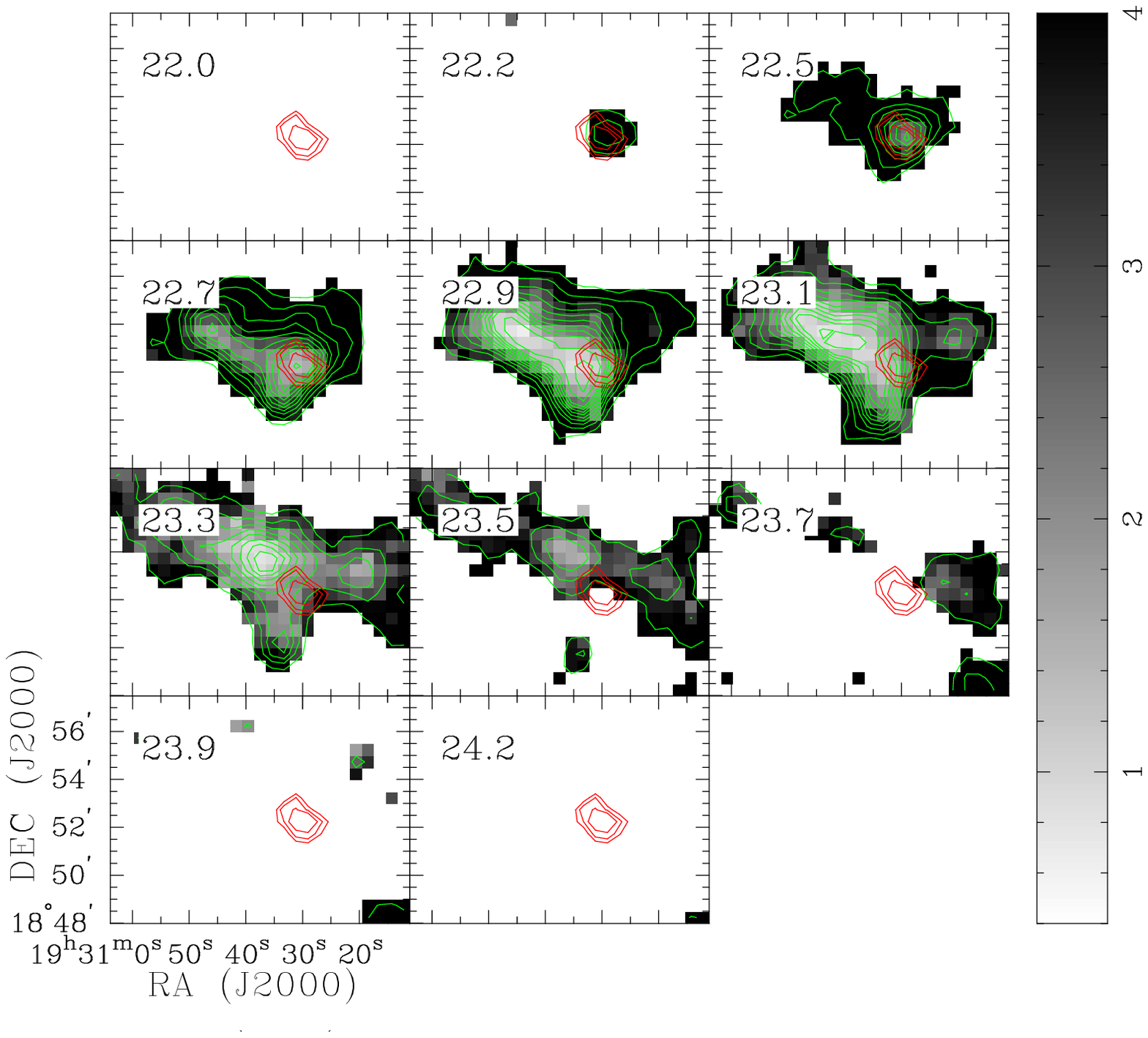} 

\caption{
$^{12}$CO(2--1)/$^{13}$CO(1--0) ratio maps (grey scale) 
corresponding to the same channels in Fig.~\ref{23_channel}.
The overlaid contours represent the \threeco intensities. The contours
increase from 0.4 K in every 0.2 K steps.     
}
\label{ratio_map}
\end{figure}

\begin{figure}[!b]
\centering
\epsfxsize=0.4\textwidth \epsfbox{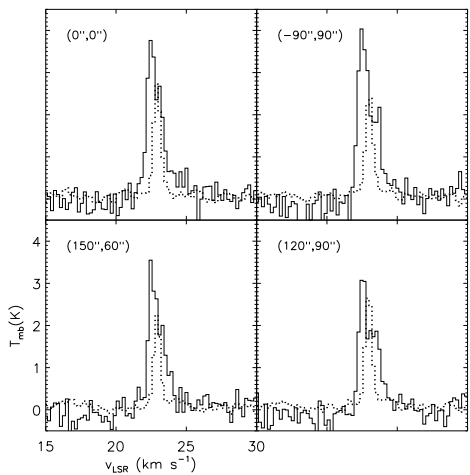} 
\caption{\twoco $J=$2--1(solid) and \threeco $J=$1--0 line profiles at positions 
where the self-absorption features are apparent. The positions 
of the spectra are marked in each frame with respect to 
(19$^{h}$30$^{m}$35.4$^{s}$,+18$^{\circ}$52$^{'}$14$^{''}$).     
}
\label{self_abs_profile}
\end{figure}

The \twoco 
emission structure of the 23 \kms cloud shown in Figure~\ref{23_channel}
is rather complicated, and 
it is helpful to look at the $^{13}$CO map first.
In Figure \ref{ratio_map}, we show 
$R_{21/10}$ ratio map
with the \threeco intensity contours overlaid. The ratios 
are derived in each channel for the pixels with $T_{mb,13}\ge 0.35$~K 
which corresponds to $5\sigma$ level. 
Note that the ratios are small where \threeco 
emission is strong. It becomes less than 1 at \threeco peaks. 
This is an indication of self-absorption 
by a relatively cold CO gas in front of warmer CO gas. 
The self-absorption is also apparent in the line profiles
where we see a dip or a plateau in the \twoco profiles at which the \threeco
intensity has maximum (Fig. \ref{self_abs_profile}).
According to the \threeco intensity contour maps, the $^{13}$CO cloud at +23 \kms 
is elongated along the northeast-southwest direction 
and is composed of several clumps. 
The clump that is spatially coincides with the PWN/IR-loop is the one in the 
southwestern end of the cloud. Its central velocity is 
22.7 \kmss, which is slightly smaller than those of the other clumps 
in the cloud. The median velocity of the 
$^{13}$CO cloud is $23.0\pm 0.2$ \kmss.

The morphology of the +23 \kms \twoco cloud is now understandable. 
At 22.5--23.1 \kmss, where the cloud is most apparent, the cloud has a 
`$\tau$' shape, 
and the \threeco emission fits into the weak inner part of this 
structure. Therefore, the weak \twoco emission in the interior
of `$\tau$' shape
should be mainly due to the absorption by the \threeco cloud.
The self-absorption can be seen either when a cold CO cloud 
with high CO opacity but low excitation temperature 
is in front of warm CO gas further away, e.g., 
a cold cloud at 1.8 kpc aligned with a warm cloud at 8.2 kpc, 
or when a CO cloud has a temperature gradient along the line of sight.
It is not unusual to see the self-absorption features in molecular clouds in the Galactic plane \citep[e.g.,][]{bieging10}. Considering the small angular sizes of the 
\twoco and \threeco clouds, the probability of a 
chance alignment of two clouds along the line of sight 
might be low. On the other hand, 
the self-absorption is prominent only to the eastern part of the \twoco cloud,
which suggests that the absorbing \threeco cloud could be distinct 
from the \twoco cloud. But this could be due to the temperature 
and density structure of the cloud. In order to quantitatively 
understand the self-absorption features, we need to solve the 
radiative transfer equation together with statistical rate equations, which is beyond the scope of this paper.

\begin{figure}[!t]
\epsfxsize=8cm \epsfbox{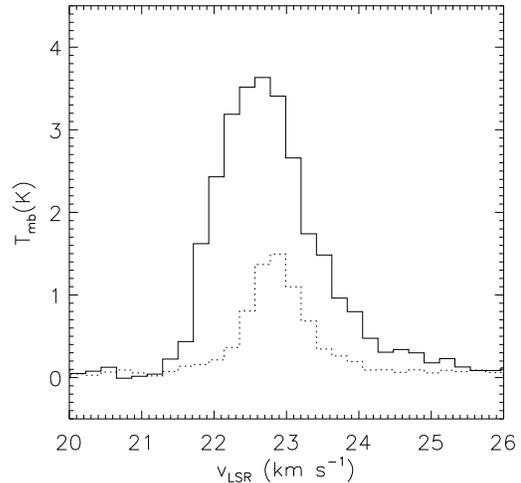} 
\caption{Average \twoco $J=$2--1 (solid) and \threeco $J=$1--0 line profiles toward 
the position of the MIR loop.
}
\label{irshell}
\end{figure}

The PWN/IR-loop spatially coincides with the southern part of 
the +23 \kms cloud. There is no obvious indication of dynamical disruption in the 
line profile (see Fig. \ref{irshell}). 
There is a wing-like structure at high
velocities (23--25 km s$^{-1}$), but this is likely due to other clouds in this area
although it needs to be confirmed with an observation of higher angular resolution. 
But in the western part of the field, there are $^{12}$CO emissions
with considerably large line widths, which are discussed in the next
section.

\subsection{Arc Cloud and Broad Line Molecular Gas at 15--30 \kmss. \label{broad_lines}}

\begin{figure}[!t]
\epsscale{1.0} \centering
\epsfxsize=0.48\textwidth \epsfbox{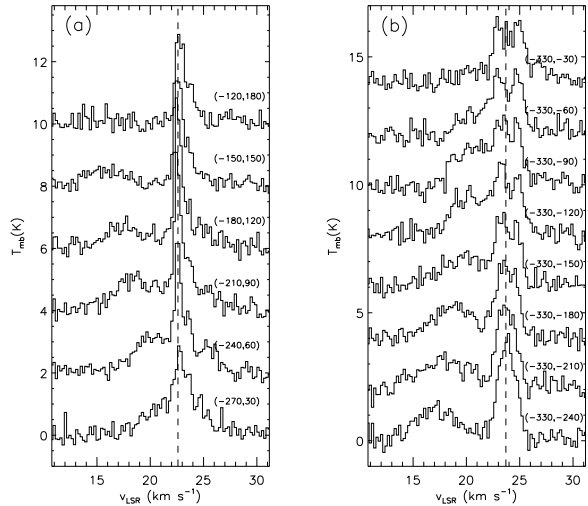}
\caption{\twoco $J$=2--1 spectra along the (a) northern and (b) southern 
parts of Arc Cloud at every 30$''$ in declination. Note the 
broad-line component shifting systematically in velocity. 
Relative coordinates in arc-seconds from the map center are indicated in parentheses. 
The position angle in (a) is 45$^\circ$ at the position of the bottom spectrum. 
Each spectrum is plotted with 2 K temperature offset. 
The vertical dashed lines are drawn just for velocity reference.    
}
\label{arc_spectrum}
\end{figure}

\begin{figure}[t]
\centering
\epsfxsize=0.35\textwidth \epsfbox{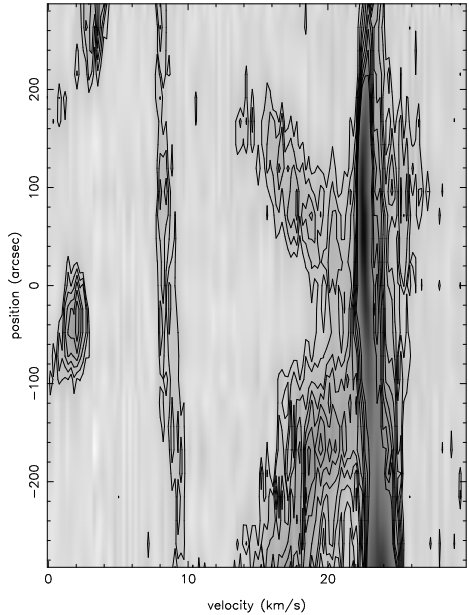}
\caption{
Position-velocity diagram at the offset position ($\Delta\alpha,\Delta\delta$)$=($-230$^{''}$,$0^{''}$) sliced in the position angle of 20 degrees. 
Contour levels increase 
from 0.4 K in 0.2 K steps. In order to show Arc Cloud clearly cutting position and angle
are chosen dissimilar to Fig.~\ref{arc_spectrum}. The southern part is brighter and 
more complex than the northern part.
}
\label{pv_diagram}
\end{figure}

A distinct structure in the observed field is Arc Cloud 
seen in Figure \ref{overall_channel_map} 
at velocities between 15 and 23 \kms 
as an elongated cloud stretching over $5'$ along
the northeast-southwest direction. 
Along the structure, 
the \twoco line profiles have large line widths and their central velocities
vary systematically as shown in Figures~\ref{arc_spectrum} 
and \ref{pv_diagram} which are the plot of spectra and 
position-velocity diagram along the arc structure.
The systematic variation of the central velocity is clear;
the velocity increases from 15 \kms at the both ends of
Arc Cloud to 23 \kms at the middle point where the northern and southern arc structure merge.
The velocity gradient is 
$\sim 8$ \kms\ over 150$''$ from the middle to the north and south ends.
At a distance of 8 kpc, it corresponds to $\sim 1.4$ \kms pc$^{-1}$.
The derived properties of Arc Cloud are summarized in Table~\ref{shell_table}.
Velocity range is chosen as 15--22 \kms to get the integrated intensity in Table~\ref{shell_table} since the main cloud, some of which may not belong to Arc Cloud, starts to appear at $> 22$ \kmss. $\rm {T_{mb}}$  less than 0.4 K (several $\sigma$ level) was clipped out in the integration. 

\begin{figure}[!ht]
\centering \epsfxsize=8cm 
\epsfbox{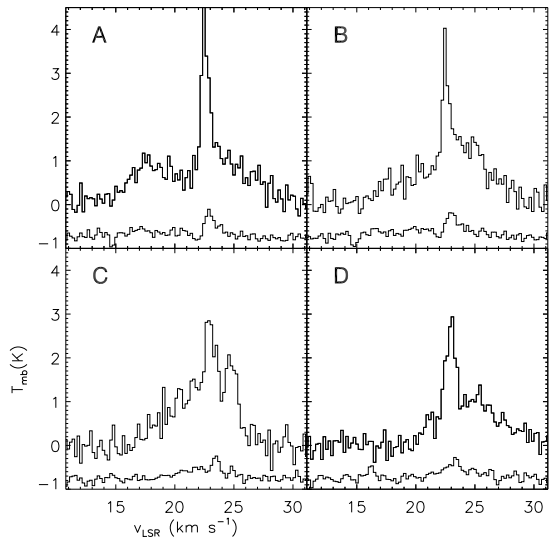} 
\caption{ 
Some representative spectra showing broad line features. 
Their positions are marked in Fig.~\ref{bml_distribution}. 
The spectra at the bottom of each frame 
are the corresponding GRS $^{13}$CO $J$=1--0 spectra with -0.8 K offsets.
}
\label{bml_spectrum}
\end{figure}

In addition to Arc Cloud, 
there are other broad lines at higher velocities in the northwestern 
area of the field. This can be seen, for example, in the line profiles of the 
northern part of Arc Cloud in Figure~\ref{arc_spectrum}, but some  
representative line profiles are shown in Figure~\ref{bml_spectrum}. 
In order to investigate the nature of the broad line emission, we 
fit the line profiles with 2 Gaussian components. 
Some profiles, i.e., the 
profiles with both the blue-shifted and red-shifted broad components, 
need three Gaussian components to describe the profiles 
correctly. But the two-component fit should be useful to see 
the basic properties of the broad components.
We used MPFIT, which is a least-squares fitting tool based on the
Levenberg-Marquardt algorithm \citep{markwardt09}.
In the fit, we leave the height, central velocity, and the width 
of individual components
free with reasonable lower and upper boundaries.

\begin{figure}[!ht]
\centerline {\includegraphics[width=0.48\textwidth]{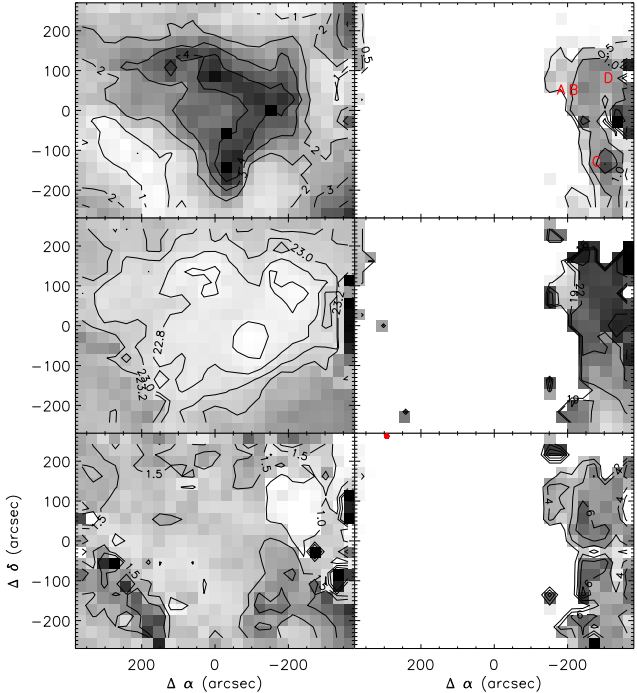}}
\caption{Distribution of the parameters of two Gaussian components 
describing the line profiles between $v_{\rm LSR}=20$ and 28 \kmss. 
The left column shows peak temperatures, central velocities, and 
line widths of narrow component while the right column shows 
the same plots for the broad lines. The positions of the spectra shown 
in Fig.~\ref{bml_spectrum} are marked in the top right panel.}     
\label{bml_distribution}
\end{figure}

\begin{figure}[!t]
\centerline {\includegraphics[width=0.48\textwidth]{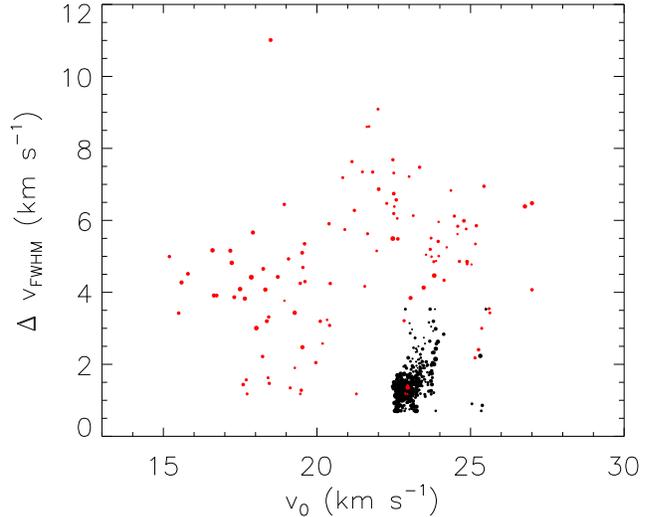}}
\caption{Line width vs. central velocity of the two Gaussian components 
in Fig. \ref{bml_distribution}.
The narrow components (black dots) are confined to a small area with  
a central velocity of 23.1$\pm$0.24 \kms and a
median width of 0.58 \kmss. 
On the other hand, the broad component(red dots) are scattered 
over a large area with a central velocity of $21.8\pm 6.2$ \kms
and a median width of 2.1 \kmss. The areas of the circles are proportional 
to their heights.}     
\label{vel.dist}
\end{figure}

Figure~\ref{bml_distribution} shows the distributions of 
the height, central velocity, and width of the two Gaussian components.
The height map of the first (narrow) component shows the morphology 
of the 23 \kms component nicely. 
The second component represents the broad-line component, and 
its maps show that the broad-line component is 
confined to the western part of the field. 
In the central velocity map, 
we can identify Arc Cloud that has 
a velocity gradient that we described above.
The broad-lines at higher velocities appear 
mainly to the western side of Arc Cloud. 
The very large width at the tip of the arc-like cloud indicates 
that, at those pixels, both blue-shifted and red-shifted 
velocity components appear (see Fig. \ref{bml_spectrum}).
Figure~\ref{vel.dist} plots the line width versus central velocity of the two 
Gaussian components. 
The black dots confined to the central velocity of 23 \kms and 
a width (FWHM) of 1.4 \kms represents the first Gaussian component, or 
the +23 \kms cloud component. The 
red dots represent the second Gaussian component, or the
broad component, and they are scattered over the plane.
The points with central velocities less than 20 \kms might be 
from Arc Cloud while the points with widths greater 
than 7--8 \kms have both red-shifted and blue-shifted 
broad components.
Figure \ref{vel.dist} shows that the broad lines 
in the western part of Arc Cloud have central velocities
of 20--26 \kms and widths of 2--7 \kmss.

\begin{table}[!t]
\begin{center}
\centering
\caption{$^{12}$CO $J=$2--1 Line Properties of Arc Cloud \label{shell_table}}
\doublerulesep2.0pt
\renewcommand\arraystretch{1.5}
\begin{tabular}{lc} 
\hline \hline
Center            & (19$^{h}$30$^{m}$19$^{s}$,+18$^{\circ}$52$^{'}$14$^{''}$)      \\
Length            & $5'$        \\
Velocity range    & 15--23 \kms \\
Velocity gradient & $\sim 3$~km s$^{-1}$ arcmin$^{-1}$ \\ 
$\Delta v$(FWHM)  &  3.6 \kms--4.5 \kms \\
Peak temperature  & $\le 1$~K \\
$\int T_{mb} dv$  & $\sim 242$ K \kms\\ \hline
\end{tabular}
\end{center}
\end{table}

\section{Discussion and Conclusion}

It has been previously proposed that the +23 \kms and +53 \kms 
molecular clouds are possibly associated with the SNR by 
\citet{koo08} and \citet{leahy08}, respectively.
They inspected the FCRAO $^{13}$CO survey data and found that 
those clouds have spatial correlation with the PWN, which
led them to propose the association although there is no 
direct evidence for the interaction. 
Our $^{12}$CO $J$=2--1 map confirms the spatial correlation.
The +23 \kms component spatially coincides with the IR shell while the 
53 \kms component appears to be in contact with the
eastern boundary of the PWN.
The kinematic distances to these clouds are 
1.8/8.2 kpc and 6.9 kpc, respectively.
The distance to G54.1+0.3 has previously been determined in several studies. 
The distance based on HI absorption spectrum is from 5 to 10 kpc 
\citep{koo08,leahy08}, and
the free-electron density/distance model \citep{cordes02} 
gives the distance of $9^{+1.0}_{-1.5}$ kpc
to the PSR J1930+1852 at the center of the PWN from its 
dispersion measure \citep{cam02}.
Recently, \citet{kim12} derived a distance of $6.0\pm 0.4$ kpc 
to the IR-excess stellar objects from a spectro-photometric study.
Therefore, the estimated distance to the SNR ranges from 5 to 9 kpc,
and in principle either of the cloud can be associated with 
the SNR.
 
The +53 \kms cloud is composed of several arc-minute-sized clumps forming 
a kink-shaped structure (Fig. \ref{53_channel}). The PWN is located in 
the middle where the cloud appears to kink. The spatial correlation is
suggestive for the interaction as pointed out by \citet{leahy08}.
But we do not see any direct evidence of the interaction;
its $^{12}$CO $J=$2--1 line profiles are well described by a single Gaussian with 
velocity widths (FWHM) of $1.5$--2.0 \kmss. 
It is difficult to imagine a molecular cloud right at the SN 
explosion site but not disrupted. Also, the PWN shows 
no indication that it had encountered a dense medium only 
in the east; its radio structure has a east-west symmetry. 
Therefore, we consider that the association of the +53 \kms
 cloud with the PWN is not likely.
There is weak ($\sim 1$ K) diffuse emission with broad lines 
at comparable velocities in the eastern area of the cloud. The 
association 
of this diffuse component with the +53 \kms cloud is uncertain. But 
since it extends beyond the radio boundary of the SNR, 
it is not likely that this component is associated with the SNR.

The 23 \kms cloud has a complex structure with pronounced 
self-absorption features.
The self-absorption could be either due to a 
foreground cold CO cloud along the line of sight or 
due to a temperature gradient in the cloud.
The entire cloud structure is large ($\sim10'$) and again
it is not likely that this cloud is associated with 
the SNR. There is a small clump in the southwestern end of the cloud
that coincides 
with the MIR loop \citep{koo08} at  $\sim 22$ \kms (see Fig. \ref{23_channel}). It is not impossible that 
this clump is spatially separate from the +23 \kms cloud, 
but there is no obvious indication of dynamical disruption in its line profile.

A distinct structure detected in this study is Arc 
Cloud in the west of the PWN
at velocities between 15 and 23 \kmss.
The cloud is $5'$ long and has a large line width.
The central velocity increases from both ends of the cloud
to the middle point systematically. 
The velocity at the end of the arc cloud is 15 \kmss, so that 
the velocity shift from the middle (+23 \kmss) is $\sim 8$ \kmss.
There are also other broad lines at higher velocities 
in the west of Arc Cloud. The association of 
this broad line emission with Arc Cloud is possible.
It is interesting to note that the radio brightness of 
the SNR is relatively faint in its western part where these
arc-like cloud and the broad-line clouds are located
\citep[see Fig. 4 of][]{lang10}.
The SNR shock of G54.1+0.3 is 
currently expanding at $\sim 2,000$ \kms and the 
density of the ambient medium is $\sim 0.2$ cm$^{-3}$ 
\citep{bocchino10}. 
If a dense molecular cloud is swept-up by this shock wave, 
the shock speed will drop by the square root of the 
density contrast. Therefore, Arc  
Cloud and the broad-line cloud could have been 
produced by the SNR shock if they were recently swept-up and 
if their densities were $\sim 10^4$~cm$^{-3}$.
However in this part of the sky,  
there is a large star-forming molecular cloud near the Galactic plane
at the same LSR velocity, but at a distance of 
2 kpc. It is possible that the cloud that we observe is part of 
this cloud system. A further study is required to investigate their 
association.

In conclusion, we could not detect any direct evidence for 
the interaction of either the +53 or +23 \kms cloud with the SNR
from our CO $J=$2--1 line observations.
Instead we have detected molecular gas with broad lines and 
with systematic velocity structure at $v_{\rm LSR}=$15--30 \kms 
to the west of the IR loop but inside the SNR.  
Their association with the SNR needs to be investigated with further observations. 

\acknowledgments{ 
We express thanks to Yong-Sun Park, SRAO people and Do-Young Byun for support during commissioning of the new receiver system, especially Hyun-Woo Kang for update on control software.
B.-C. K. and J.-E. L. have been supported by Basic Science
Research Program NRF-2011-00007223 and 2012-0002330, respectively, 
through the National Research Foundation of Korea (NRF) funded by
the Ministry of Education, Science and Technology.
This publication makes use of molecular line data from the Boston University-FCRAO Galactic Ring Survey (GRS). }

\end{document}